\documentclass{article}

\pdfoutput=1

\usepackage{amsmath}
\usepackage{amsfonts}
\usepackage{amssymb}
\usepackage{latexsym}
\usepackage{ifthen}
\usepackage{theorem}
\usepackage{graphicx}
\usepackage{array}
\usepackage{caption,subcaption}

\newtheorem{prop}{Proposition} 
\newtheorem{lemma}[prop]{Lemma}
\newtheorem{theorem}[prop]{Theorem}
\newtheorem{cor}[prop]{Corollary}

\title{Order distances and split systems}

\date{\today}

\author{Vincent Moulton\\University of East Anglia, School of Computing Sciences\\ Norwich, NR4 7TJ, UK\\v.moulton@uea.ac.uk
\and Andreas Spillner\\Merseburg University of Applied Sciences\\ 06217 Merseburg, Germany\\andreas.spillner@hs-merseburg.de}

\begin{document}

\maketitle

\begin{abstract}
Given a distance \(D\) on a finite set \(X\) with \(n\) elements,
it is interesting to understand how the ranking \(R_x = z_1,z_2,\dots,z_n\) 
obtained by ordering the elements in \(X\) according to increasing
distance \(D(x,z_i)\) from \(x\), varies with different choices of $x \in X$.
The \emph{order distance} $O_{p,q}(D)$ is a distance on $X$
associated to $D$
which quantifies these variations, 
where \(q \geq \frac{p}{2} > 0\) are parameters that
control how ties in the rankings are handled.
The order distance $O_{p,q}(D)$ of a distance $D$ has been intensively 
studied in case $D$ is a \emph{treelike} distance (that is, $D$ arises
as the shortest path distances in an edge-weighted tree 
with leaves labeled by $X$), but relatively little is known
about properties of $O_{p,q}(D)$ for general~\(D\). 
In this paper we study the order distance  
for various types of distances that
naturally generalize treelike distances in that 
they can be generated by  \emph{split systems}, i.e. 
they are examples of so-called \emph{$l_1$-distances}. In particular 
we show how and to what extent properties of the split systems
associated to the distances $D$ that we study can be used to 
infer properties of~$O_{p,q}(D)$.
\end{abstract}

  
\section{Introduction}
\label{section:introduction}

A \emph{distance} $D$ on a finite, non-empty set~\(X\) 
is a symmetric map \(D : X \times X \rightarrow \mathbb{R}\)
with \(D(x,x)=0\) and \(D(x,y) \geq 0\) for all \(x,y \in X\).
Following \cite{bon-gue-96a} we 
associate a new distance $O_{p,q}(D)$ on $X$ to the 
distance $D$ for 
\(p,q \in \mathbb{R}\) with \(q \geq \frac{p}{2} > 0\) as follows. 
For each $u,v \in X$ with \(u \neq v\)  we define the sets
$$
X_{u,v} = \{x \in X : D(u,x) < D(v,x)\},
$$ 
and 
$$
E_{\{u,v\}} = \{x \in X : D(u,x) = D(v,x)\}.
$$ 
Note that the sets \(X_{u,v}\), \(X_{v,u}\) and \(E_{\{u,v\}}\) are
pairwise disjoint and that their union is \(X\).  Now, for any bipartition or \emph{split} $\{A,B\}$
of $X$, let \(D_{\{A,B\}}\) be the distance on \(X\) given by taking 
\(D_{\{A,B\}}(x,y) = 1\) if \(|A \cap \{x,y\}| = |B \cap \{x,y\}| = 1\) 
and \(D_{\{A,B\}}(x,y) = 0\) otherwise for all \(x,y \in X\).
The  \emph{order distance} $O_{p,q}(D)$  associated to $D$ is then defined as
\begin{align}
\label{equ:general:formula:od:splits}
O_{p,q}(D) = &\sum_{\substack{(u,v) \in X \times X, u \neq v\\ \emptyset \subsetneq X_{u,v} \subsetneq X}} \frac{p}{2} \cdot D_{\{X_{u,v},X-X_{u,v}\}} \notag\\
            &\quad \quad + \sum_{\substack{\{u,v\} \in \binom{X}{2}\\ \emptyset \subsetneq E_{\{u,v\}} \subsetneq X}} (q-\frac{p}{2}) \cdot D_{\{E_{\{u,v\}},X-E_{\{u,v\}}\}},
\end{align} 
where \(\binom{X}{2}\) denotes the set of all 2-element subsets of~\(X\).

The order distance $O_{p,q}(D)(x,y)$ can be regarded as the amount 
by which the two rankings \(z_1,z_2,\dots,z_n\) 
and \(z_1',z_2',\dots,z_n'\) of the elements in \(X\) generated
by ordering these elements according to increasing distances \(D(x,z_i)\) from \(x\)
and \(D(y,z_i')\) from \(y\), respectively, differ \cite{bon-gue-96a}.
Note that, as pointed out in \cite{bon-gue-96a}, to ensure that \(O_{p,q}(D)\) satisfies the triangle inequality 
we must require \(q \geq \frac{p}{2}\), and that, 
for any \(c \in \mathbb{R}\) with \(c > 0\), \(O_{c \cdot p, c \cdot q}(D) = c \cdot O_{p,q}(D)\).

\begin{figure}
\centering
\includegraphics[scale=1.0]{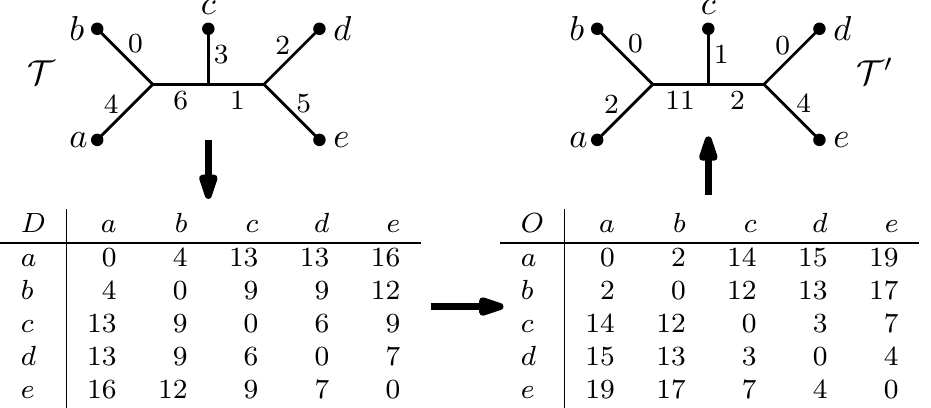}
\caption{A tree \(\mathcal{T}\) with non-negative edge weights whose leaves are labeled
	by the elements in the set \(X=\{a,b,c,d,e\}\) and the
	distance \(D\) of shortest path distances between the leaves of \(\mathcal{T}\).
	The associated order distance \(O=O_{2,1}(D)\) can also be represented by the tree
	\(\mathcal{T}\) by adjusting the weights of its edges. Note that 
	each edge of \(\mathcal{T}\) corresponds to a split of $X$.}
\label{figure:example:treelike:distance}
\end{figure}

Most previous work concerning order distances 
has focused on their properties for \emph{treelike} distances, that is, distances that
arise by taking lengths of shortest paths between pairs of leaves of a tree
(see e.g. \cite{bon-gue-96a,gue-97b,gue-97a,gue-98a}). 
In Figure~\ref{figure:example:treelike:distance} we present an example
of the order distance~\(O=O_{p,q}(D)\) with \(p=2\) and \(q=1\) associated to a 
treelike distance~\(D\). Note that in this example the 
order distance  \(O=O_{p,q}(D)\) is also treelike since, as we can see in
the figure, it can be represented by  
adjusting the weights of the edges in~\(\mathcal{T}\).
In fact, this is no coincidence: The main result of Bonnot et al. in \cite{bon-gue-96a}
establishes that if \(D\) is a treelike distance which can be 
represented by giving non-negative weights to the edges 
in some tree, then for $q = \frac{p}{2}$, after possibly adjusting the edge weights, the order distance
\(O_{p,q}(D)\) can also be represented by the same tree.

In this paper, we aim to better understand to what extent this
result can be extended to more general distances.
In particular we shall focus on order distances for 
so-called \(\ell_1\)-\emph{distances} (cf. \cite[Ch. 4]{dez-97}), that is,
distances \(D\) on \(X\) for which there exists a set \(\mathcal{S}\) of splits of \(X\) or \emph{split system},
together with a non-negative weighting \(\omega: \mathcal{S} \rightarrow \mathbb{R}\) such that
\[D = \sum_{S \in \mathcal{S}} \omega(S) \cdot D_S,\]
also referred to as an \(\l_1\)-\emph{decomposition} of \(D\).
It is natural to consider \(\ell_1\)-distances in the context of order distances
since it follows directly from 
Equation~(\ref{equ:general:formula:od:splits})
that the order distance \(O=O_{p,q}(D)\) for \emph{any} distance $D$ 
is an \(\ell_1\)-distance. 

Interestingly, Bonnot et al.'s result can be re-expressed in \(\ell_1\)-terminology
as follows. To any tree \(\mathcal{T}\) with leaves labeled by $X$ we can associate 
the split system \(\mathcal{S}= \mathcal{S}_\mathcal{T}\) consisting of the splits of \(X\) that 
correspond to the edges of the tree  (e.g. the
edge in \(\mathcal{T}\) in Figure~\ref{figure:example:treelike:distance}
with weight 6 corresponds to the split $\{\{a,b\},\{c,d,e\}\}$). Bonnot et al.'s result 
then states that in case $q = \frac{p}{2}$,  for every non-negative  weighting \(\omega\) of 
the splits in \(\mathcal S\), there exists a non-negative weighting \(\omega'\) of 
the splits in \(\mathcal{S}\) such that
\begin{equation}
\label{equation:orderly:general}
O_{p,q}(\sum_{\{A,B\} \in  \mathcal S} \omega(S) \cdot D_{\{A,B\}}) = \sum_{\{A,B\} \in  \mathcal S} \omega'(S) \cdot D_{\{A,B\}}.
\end{equation}
Since the split system \(\mathcal S\) arises from a tree, it has a special 
combinatorial property known as \emph{compatibility} (see Section~\ref{subsec:treelike:distances:compatible:ss}). 
In this paper, we will explore under what conditions 
Equation~(\ref{equation:orderly:general}) might 
hold for other split systems \(\mathcal{S}\) that are not necessarily compatible.

The rest of the paper is structured as follows.
In Section~\ref{subsec:treelike:distances:compatible:ss} we present
some preliminaries concerning the relationship between
treelike distances and compatible split systems.
Then, in Section~\ref{sec:treelike:revisited}, we prove a variant of Bonnot et al.'s
result for arbitrary  \(q \geq \frac{p}{2} > 0\) in the special case where 
the split system underlying a tree is maximal (Theorem~\ref{theo:compatible:general:pq:order:dist}). 
In Section~\ref{section:split:system:sd} we
focus on the split system associated to a distance $D$ on $X$ 
that forms the index set for the first sum in 
Equation~(\ref{equ:general:formula:od:splits}). In particular, we
give a tight upper bound on its size (Theorem~\ref{theo:upper:bound:size:sd}), and also
a characterization for when it is compatible (Theorem~\ref{theorem:six:point:sd:compatible}).

In Section~\ref{section:circular:split:systems}
we introduce the concept of an \emph{orderly} split system. 
These are essentially split systems for which Equation~(\ref{equation:orderly:general})
holds in case $p=\frac{q}{2}$. Compatible split systems are special examples of orderly
split systems, and we show that the more general \emph{circular} split systems \cite{ban-dre-92a}
also enjoy this property in case they have maximum size (Theorem~\ref{theo:circ:split:system:orderly}).
In Sections~\ref{section:max:lin:independent}
and \ref{section:flat:split:systems} we then explore to what extent
this latter result can be extended to the even more general class of so-called \emph{flat}
split systems \cite{bry-dre-07a,spi-ngu-12a}. In particular, we show that within
the class of maximum sized flat split systems, the orderly
split systems are precisely those that are circular (Theorem~\ref{flat-orderly}).
In Section~\ref{section:algorithms:computing:order:distance} we briefly
look into consequences of our results on efficiently computing order distances.
We conclude in Section~\ref{section:comcluding remarks} with some 
possible directions for future work.

\section{Preliminaries}
\label{subsec:treelike:distances:compatible:ss}

For the rest of this paper \(X\) will denote a finite non-empty set with $|X|=n$, and 
\(\mathcal{S}(X)\) the set consisting of all possible splits of \(X\).
We also use \(A|B = B|A\) to denote a split \(\{A,B\}\) of \(X\) into two
non-empty subsets \(A\) and \(B\).
We call any non-empty subset \(\mathcal{S} \subseteq \mathcal{S}(X)\)
a \emph{split system on}~\(X\). A pair \((\mathcal{S},\omega)\) consisting
of a split system \(\mathcal{S}\) on \(X\) and a weighting \(\omega: \mathcal{S} \rightarrow \mathbb{R}_{\ge 0}\)
is called a \emph{weighted split system on}~\(X\) and we denote
by \(D_{(\mathcal{S},\omega)} = \sum_{S \in \mathcal{S}} \omega(S) \cdot D_S\)
the distance \emph{generated} by the weighted split system \((\mathcal{S},\omega)\).
We emphasize that throughout this paper \emph{the weights of the splits in a weighted split systems will
always be non-negative}. 

A split system \(\mathcal{S}\) on \(X\) is \emph{compatible} 
if, for any two splits \(A_1|B_1\) and \(A_2|B_2\) in \(\mathcal{S}\), at least one of the
intersections \(A_1 \cap A_2, \ A_1 \cap B_2, \ B_1 \cap A_2 \ \text{and} \ B_1 \cap B_2\) is empty.
The splits in a compatible split system 
\(\mathcal{S}\) on \(X\) are in one-to-one correspondence with the edges of a (necessarily)
unique \(X\)-\emph{tree} \(\mathcal{T} = (V,E,\varphi)\),
that is, a graph theoretic tree \(\mathcal{T} = (V,E)\) with vertex set \(V\) and 
edge set \(E \subseteq \binom{V}{2}\) together with a map \(\varphi: X \rightarrow V\)
such that the full image \(\varphi(X)\) contains all vertices of degree at most two.
We denote the compatible split system represented by the edges of an \(X\)-tree \(\mathcal{T}\)
by \(\mathcal{S}_{\mathcal{T}}\).
The edges of the \(X\)-tree \(\mathcal{T}\) in Figure~\ref{figure:example:treelike:distance},
for example, yield the following collection of splits of \(X=\{a,b,c,d,e\}\):
\(\{a\}|\{b,c,d,e\}\), \(\{b\}|\{a,c,d,e\}\), \(\{c\}|\{a,b,d,e\}\), \(\{d\}|\{a,b,c,e\}\), \(\{e\}|\{a,b,c,d\}\),
\(\{a,b\}|\{c,d,e\}\) and \(\{a,b,c\}|\{d,e\}\) (which can be visualized 
by removal of each edge from \(\mathcal{T}\)).
Assigning to each of these splits as its weight the weight of the corresponding edge in \(\mathcal{T}\)
yields a weighted compatible split system that generates~\(D\).

Note that a compatible split system \(\mathcal{S}\)
on \(X\) is \emph{maximal}, that is, adding any further
split to \(\mathcal{S}\) yields a split system
that is no longer compatible, precisely in case it contains 
\(2n - 3\) splits, in which case it 
corresponds to a binary \(X\)-tree where the elements of \(X\)
are in one-to-one correspondence with the leaves of the tree.
Hence, maximal compatible split systems are precisely 
the maximum-sized or \emph{maximum}, for short, compatible 
split systems on \(X\).

In proofs we will make use of these facts concerning 
compatible split systems and will sometimes switch
between a weighted compatible split system and its equivalent unique representation
as an edge-weighted \(X\)-tree. Full details concerning
this correspondence can be found in~\cite{sem-ste-03a}.

\section{Treelike distances revisited}
\label{sec:treelike:revisited}

In this section, we consider properties of the order distance 
\(O_{p,q}(D)\) of a treelike distance \(D\) for arbitrary values \(q \geq \frac{p}{2} > 0\).
Note that most previous results for treelike distances, such as those in 
mentioned in the introduction, focus mainly on the case \(q = \frac{p}{2}\).

We first consider the case where \(D\) is generated by a maximum compatible split system.

\begin{theorem}
	\label{theo:compatible:general:pq:order:dist}
	For any maximum compatible
	split system \((\mathcal{S},\omega)\) on \(X\) with strictly positive weighting \(\omega\),
	the order distance \(O=O_{p,q}(D)\)
	associated to \(D = D_{(\mathcal{S},\omega)}\) can be expressed as
	\(O=D_{(\mathcal{S},\omega')}\) for some non-negative weighting \(\omega'\).
\end{theorem}

\emph{Proof:}
Let \((\mathcal{S},\omega)\) be a maximum compatible split system 
on \(X\) with strictly positive weighting \(\omega\).
First note that in \cite{kea-97a} it is shown that if \(\mathcal{S}\) is compatible then
the splits \(X_{u,v}|X-X_{u,v}\) and \(X_{v,u}|X-X_{v,u}\) are contained in \(\mathcal{S}\) for
all \((u,v) \in X \times X\) with \(\emptyset \subsetneq X_{u,v} \subsetneq X\).
Moreover, in view of the fact that maximum compatible split systems 
with strictly positive weighting are
precisely those that can be represented by binary \(X\)-trees with strictly positive edge weights
whose leaves are in one-to-one correspondence with the elements in~\(X\),
we must have \(D(u,v) > 0\) for all \(u,v \in X\) with \(u \neq v\)
and, for any \(\{u,v\} \in \binom{X}{2}\) with \(\emptyset \subsetneq E_{\{u,v\}} \subsetneq X\),
the splits \(X_{u,v}|X-X_{u,v}\), \(X_{v,u}|X-X_{v,u}\) and \(E_{\{u,v\}}|X - E_{\{u,v\}}\) correspond
to three edges that share a single vertex in the binary \(X\)-tree that represents \(\mathcal{S}\).
In particular, the split \(E_{\{u,v\}}|X - E_{\{u,v\}}\) must be contained in \(\mathcal{S}\).
Hence, from Equation~(\ref{equ:general:formula:od:splits}) and in view of the
assumption \(q \geq \frac{p}{2}\) we have
\begin{align*}
O_{p,q}(D)
&= \sum_{\substack{(u,v) \in X \times X, u \neq v\\ \emptyset \subsetneq X_{u,v} \subsetneq X}} \frac{p}{2} \cdot D_{X_{u,v}|X-X_{u,v}}\\
&\quad \quad + \sum_{\substack{\{u,v\} \in \binom{X}{2}\\ \emptyset \subsetneq E_{\{u,v\}} \subsetneq X}} (q-\frac{p}{2}) \cdot D_{E_{\{u,v\}}|X-E_{\{u,v\}}}\\
&= \sum_{S \in \mathcal{S}} \omega'(S) \cdot D_{S}
\end{align*}
for some suitable non-negative weighting \(\omega'\) of the splits in \(\mathcal{S}\), as required.
\hfill\(\blacksquare\)\\

Note that the assumption in Theorem~\ref{theo:compatible:general:pq:order:dist}
that the compatible split system is maximum is necessary.
Indeed, in \cite[p. 258]{bon-gue-96a}, an example is presented
which provides a weighted, non-maximum compatible split system $\mathcal{S}$ on $X$, $|X|=4$, 
such that for any \(q > \frac{p}{2}\), the order distance \(O=O_{p,q}(D)\) 
associated to \(D = D_{(\mathcal{S},\omega)}\) cannot be expressed as
\(O=D_{(\mathcal{S},\omega')}\) for any non-negative weighting \(\omega'\) of the 
splits in \(\mathcal{S}\). There exists, however, a compatible
superset \(\mathcal{S}'' \supseteq \mathcal{S}\) of splits such that
\(O=D_{(\mathcal{S}'',\omega'')}\) for some non-negative weighting \(\omega''\) in this example.

In general, even expressing the order distance by a suitable
superset of splits that belong to the same class of split systems (here compatible)
as the split system that generates \(D\) requires, in general, that we restrict to \(q = \frac{p}{2}\). 
To illustrate this in the following example, we make use 
of the fact (see e.g. \cite{sem-ste-03a})
that treelike distances \(D\) on a set \(X\) are characterized by the following 4-point condition:
For all \(a,b,c,d \in X\)
\begin{equation}
\label{equ:treelike:4:point:condition}
D(a,b) + D(c,d) \leq \max (D(a,c) + D(b,d), D(a,d) + D(b,c))
\end{equation}
must hold. Now, consider the distance \(D=D_{(\mathcal{S},\omega)}\) generated by the weighted
non-maximum compatible split system \((\mathcal{S},\omega)\) with
\begin{align*}
&\omega(\{b\}|\{a,c,d,e\}) = \omega(\{e\}|\{a,b,c,d\}) = 2, \ \omega(\{a\}|\{b,c,d,e\}) = 4 \ \text{and}\\
&\omega(\{c\}|\{a,b,d,e\}) = \omega(\{d\}|\{a,b,c,e\}) = \omega(\{c,d\}|\{a,b,e\}) = 1
\end{align*}
on the 5-element set \(X=\{a,b,c,d,e\}\). 
We obtain the order distance \(O=O_{p,q}(D)\) with
\begin{align*}
&O(a,b) = O(a,e) = 4p + 3q, \ O(a,c) = 4p + 5q\\
&O(b,e) = p + 4q \ \text{and} \ O(c,e) = O(b,c) = 2p + 4q.
\end{align*}
In view of Equation~(\ref{equ:treelike:4:point:condition}), a weighted compatible split system
\((\mathcal{S}',\omega')\) with \(O=D_{(\mathcal{S}',\omega')}\) and \(\mathcal{S} \subseteq \mathcal{S}'\)
can exist only if
\[O(a,c)+O(b,e) = 5p + 9q \leq 6p + 7q = \max(O(a,b)+O(c,e),O(a,e)+O(b,c))\]
holds. But this implies \(q \leq \frac{p}{2}\) and, thus \(q = \frac{p}{2}\).

Note that the distance \(D\) in the previous example
is not only treelike but even an \emph{ultrametric}, 
that is, \(D(x,z) \leq \max(D(x,y),D(y,z))\) holds for all \(x,y,z \in X\).
In \cite{gue-97b} it is shown that for 
any ultrametric \(D\) generated by a weighted compatible split system
\((\mathcal{S},\omega)\) with strictly positive weighting \(\omega\),
the associated order distance
\(O=O_{p,\frac{p}{2}}(D)\) can be expressed as \(O=D_{(\mathcal{S},\omega')}\)
for some strictly positive weighting \(\omega'\) of the splits in \(\mathcal{S}\),
that is, \emph{all} splits in \(\mathcal{S}\) are used to generate~\(O\).
As pointed out in \cite{gue-98a}, however, this property does not characterize
ultrametrics: there are examples of distances \(D\) that have this property but 
are not ultrametrics. Moreover, it is shown in \cite{gue-97b} that 
the order distance of an ultrametric is, in general, not an ultrametric.

\section{The midpath split system of a distance}
\label{section:split:system:sd}

Given a distance \(D\) on $X$ we define the \emph{midpath split system}
\(\mathcal{S}_D\) associated to $D$ to be the set of splits of $X$ of the
form $S_{u,v}=X_{u,v}|X - X_{u,v}$ for $u,v \in X$ 
with \(u \neq v\) and $\emptyset \subsetneq X_{u,v} \subsetneq X$.
Note that the splits in \(\mathcal{S}_D\) are precisely those 
appearing in the index set of the first sum in Equation~(\ref{equ:general:formula:od:splits}).
We chose the name for \(\mathcal{S}_D\) since it is closely related to the 
\emph{midpath phylogeny} introduced in \cite{kea-98a}.
In this section, we consider general properties of the 
split system \(\mathcal{S}_D\), including a  
characterization in terms of $D$ for when this split system is compatible.

First note that, as a direct consequence of the definition of \(\mathcal{S}_D\), it follows
that \(\mathcal{S}_D\) contains at most \(n(n-1)\) splits.
As \(|\mathcal{S}(X)| = 2^{n-1}-1\), we immediately see
that for small \(n\) this upper bound is not tight for \(|\mathcal{S}_D|\). 
Nevertheless we have the following result:

\begin{theorem}
\label{theo:upper:bound:size:sd}
Let \(D\) be a distance on a set \(X\) with \(n \ge 1\) elements.
Then we have \(|\mathcal{S}_D| \leq n(n-1)\) and,
for all sufficiently large \(n \in \mathbb{N}\), this bound
is tight.
\end{theorem}

\emph{Proof:}
It remains to show that the upper bound \(n(n-1)\) is tight for
all sufficiently large \(n\). To this end, consider a distance
\(D\) on  \(X\) such that, for all \(\{u,v\} \in \binom{X}{2}\),
the value \(D(u,v) = D(v,u)\) is selected randomly from the set \(\{1,2\}\),
with both values having the same probability of being selected. 
We now argue that, for sufficiently large \(n\), the probability that
\(|\mathcal{S}_D| = n(n-1)\) is strictly greater than~0.

Note that \(D\) satisfies the triangle inequality and
that \(D(x,y) > 0\) for all \(\{x,y\} \in \binom{X}{2}\),
implying that the splits \(S_{x,y}\) and \(S_{y,x}\) exist.
Moreover, in order to have \(|\mathcal{S}_D| = n(n-1)\), 
for any two distinct \(\{u,v\}\), \(\{a,b\} \in \binom{X}{2}\),
the splits \(S_{u,v}\), \(S_{v,u}\), \(S_{a,b}\) and \(S_{b,a}\)
must be pairwise distinct. 
Now, it follows immediately from the definition of \(D\) that the
probability of \(S_{u,v} = S_{v,u}\) is \(\frac{4}{2^n}\).
More generally, the probability that
at least two of the splits \(S_{u,v}\), \(S_{v,u}\), \(S_{a,b}\) and \(S_{b,a}\) 
coincide is bounded by \(d \cdot c^n\) for some constants
\(0 < d\) and \(0 < c < 1\). This implies that the probability of
\(|\mathcal{S}_D| < n(n-1)\) is at most
\[\binom{n}{2} \cdot \left ( \binom{n}{2} -1 \right ) \cdot d \cdot c^n,\]
which is strictly less than~1 for sufficiently large~\(n\), as required.
\hfill\(\blacksquare\)\\

The remainder of this section is devoted to giving a characterization 
of those distances \(D\)  for which \(\mathcal{S}_D\) is compatible.
In \cite{kea-97a}, a 6-point condition is given
that characterizes for a distance \(D\)  when
\begin{itemize}
\item[(i)]
the split system \(\mathcal{S}_D\) is compatible and 
\item[(ii)]
there are no \(a,b,c,d \in X\) with
\(D(a,b) > 0\), \(D(c,d) > 0\) and
\(X_{a,b} \subsetneq X_{c,d} \subsetneq X - X_{b,a}\).
\end{itemize}
Note that for \emph{generic} distances~\(D\), that is,
\(D(a,b) > 0\) and \(D(a,b) \neq D(a',b')\)
holds for all \(\{a,b\},\{a',b'\} \in \binom{X}{2}\) with \(\{a,b\} \neq \{a',b'\}\),
the aforementioned 6-point condition characterizes when 
\(\mathcal{S}_D\) is compatible, 
because in this case condition (ii) cannot
be violated in view of the fact that for a generic distance 
\(X_{v,u} = X - X_{u,v}\) holds for all \(\{u,v\} \in \binom{X}{2}\).

\begin{figure}
\centering
\includegraphics[scale=0.95]{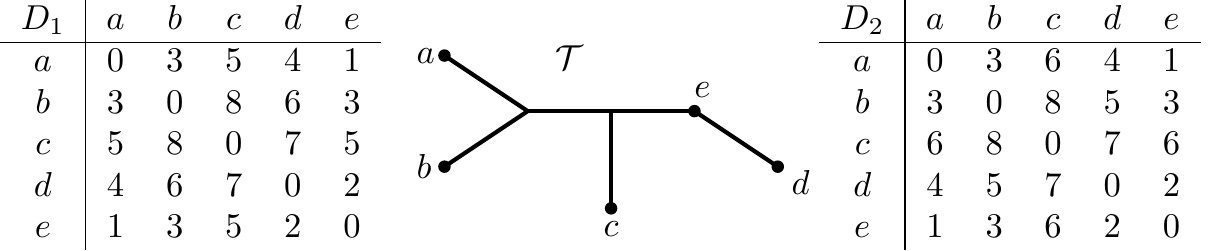}
\caption{A distance \(D_1\) on \(X=\{a,b,c,d,e\}\) such that \(\mathcal{S}_{D_1}\) is compatible
but \(D_1\) has no midpath phylogeny. The edges of the \(X\)-tree \(\mathcal{T}\)
represent the splits in \(\mathcal{S}_{D_1}\). For the distance \(D_2\)
we obtain \(\mathcal{S}_{D_2} = \mathcal{S}_{D_1}\) but \(D_1\) and \(D_2\)
do not yield the same ranking of \(\binom{X}{2}\).}
\label{figure:counterexample:no:midpath}
\end{figure}

So, we start by providing in Figure~\ref{figure:counterexample:no:midpath}
an example of a non-generic distance \(D_1\) 
for which \(\mathcal{S}_{D_1}\) is compatible but condition (ii) is violated.
More precisely we have 
\[X_{a,e} = \{a\} \subsetneq X_{b,d} = \{a,b\} \subsetneq X - X_{e,a} = \{a,b,c\}.\]
This example also illustrates the meaning of condition (ii) in terms of the
\(X\)-tree \(\mathcal{T}\) representing the splits in \(\mathcal{S}_{D_1}\):
If the edges representing the splits \(S_{a,e}\) and \(S_{e,a}\) do not coincide
they are required to share a vertex. The edges representing the
splits \(S_{a,e} = \{a\}|\{b,c,d,e\}\) and \(S_{e,a} = \{a,b,c\}|\{d,e\}\) in our example,
however, do not share a vertex.
This implies that the 6-point condition
given in \cite{kea-97a} does not provide the 
characterization for arbitrary distances that we are looking for.
Another aspect illustrated in Figure~\ref{figure:counterexample:no:midpath}
is that distances \(D_1\) and \(D_2\) on the same set \(X\)
with \(\mathcal{S}_{D_1} = \mathcal{S}_{D_2}\) do not necessarily
yield the same ranking of \(\binom{X}{2}\), not even if \(D_1\) and \(D_2\) yield the same set \(X_{u,v}\) 
for all \(u,v \in X\).

Our characterization for when \(\mathcal{S}_D\) is compatible will
also be a 6-point condition. Actually, it is more convenient to state and prove a
characterization for when \(\mathcal{S}_D\) is \emph{not} compatible. The structure of
the proof of Theorem~\ref{theorem:six:point:sd:compatible} is similar
to the proof of the 6-point condition in \cite{kea-97a}.

\begin{theorem}
\label{theorem:six:point:sd:compatible}
Let \(D\) be a distance on a set \(X\). The split system
\(\mathcal{S}_D\) is not compatible if and only if 
there exist \(a,b,s,t,x,y \in X\) with
\(a \neq b\), \(s \neq t\), \(x \neq y\) such that
one of the following holds:
\begin{itemize}
\item[(1)]
\(D(x,a) < D(y,a)\) and
\(D(y,b) \leq D(x,b)\) and either
\(D(a,s) < D(a,t)\), \(D(b,s) < D(b,t)\), \(D(x,t) \leq D(x,s)\) and \(D(y,t) \leq D(y,s)\) or
\(D(a,s) \leq D(a,t)\), \(D(b,s) \leq D(b,t)\), \(D(x,t) < D(x,s)\) and \(D(y,t) < D(y,s)\)
\item[(2)]
\(D(x,a) < D(y,a)\) and
\(D(y,b) \leq D(x,b)\) and either
\(D(b,s) < D(b,t)\), \(D(a,t) \leq D(a,s)\), \(D(x,s) < D(x,t)\) and \(D(y,t) \leq D(y,s)\) or
\(D(b,s) \leq D(b,t)\), \(D(a,t) < D(a,s)\), \(D(x,s) \leq D(x,t)\) and \(D(y,t) < D(y,s)\)
\end{itemize}
\end{theorem}

\emph{Proof:}
First assume that \(D\) satisfies (1) or (2). Then
the split \(S_{x,y}\) as well as one of the splits \(S_{s,t}\) or \(S_{t,s}\)
are contained in \(\mathcal{S}_D\).
Now, if \(D\) satisfies~(1) and \(S_{s,t} \in \mathcal{S}_D\),
then \(S_{x,y}\) and \(S_{s,t}\) are not compatible 
in view of \(a \in X_{x,y} \cap X_{s,t}\), \(b \in (X-X_{x,y}) \cap X_{s,t}\),
\(x \in X_{x,y} \cap (X-X_{s,t})\), \(y \in (X-X_{x,y}) \cap (X-X_{s,t})\) and,
if \(D\) satisfies~(1) and \(S_{t,s} \in \mathcal{S}_D\),
then \(S_{x,y}\) and \(S_{t,s}\) are not compatible 
in view of \(a \in X_{x,y} \cap (X-X_{t,s})\), \(b \in (X-X_{x,y}) \cap (X-X_{t,s})\),
\(x \in X_{x,y} \cap (X_{t,s})\), \(y \in (X-X_{x,y}) \cap X_{t,s}\).

Similarly, if \(D\) satisfies~(2) and \(S_{s,t} \in \mathcal{S}_D\),
then \(S_{x,y}\) and \(S_{s,t}\) are not compatible in view
of \(x \in X_{x,y} \cap X_{s,t}\), \(b \in (X-X_{x,y}) \cap X_{s,t}\),
\(a \in X_{x,y} \cap (X-X_{s,t})\), \(y \in (X-X_{x,y}) \cap (X-X_{s,t})\) and,
if \(D\) satisfies~(2) and \(S_{t,s} \in \mathcal{S}_D\),
then \(S_{x,y}\) and \(S_{t,s}\) are not compatible in view
of \(x \in X_{x,y} \cap (X-X_{t,s})\), \(b \in (X-X_{x,y}) \cap (X-X_{t,s})\),
\(a \in X_{x,y} \cap X_{t,s}\), \(y \in (X-X_{x,y}) \cap X_{t,s}\).
Hence, \(\mathcal{S}_D\) is not compatible, as required. 

Now assume that \(\mathcal{S}_D\) is not compatible. Let 
\[\mathcal{P} = \{(u,v) \in X \times X: u \neq v, \ \emptyset \subsetneq X_{u,v} \subsetneq X\}\]
be the set of those pairs of \(u,v \in X\), \(u \neq v\), with \(S_{u,v} \in \mathcal{S}_D\).
Order the pairs in \(\mathcal{P}\) arbitrarily and let
\((u_1,v_1),(u_2,v_2),\dots,(x_m,v_m)\) denote the resulting sequence.
Put \(S_i = S_{u_i,v_i}\) and \(\mathcal{S}_i = \{S_1,S_2,\dots,S_i\}\) 
for all \(1 \leq i \leq m\). Note that the split system \(\mathcal{S}_1\)
is compatible.

\begin{figure}
\centering
\includegraphics[scale=0.95]{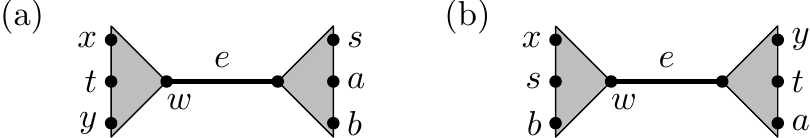}
\caption{The two possible configurations in the \(X\)-tree \(\mathcal{T}\)
referred to in the proof of Theorem~\ref{theorem:six:point:sd:compatible}.
The gray triangles indicate the two subtrees of \(\mathcal{T}\) connected to the
endpoints of the edge \(e\), one of them being the vertex~\(w\)
that lies on the path from \(x\) to \(y\). The configuration in (a) yields condition~(1)
and the configuration in (b) yields condition~(2) stated in
Theorem~\ref{theorem:six:point:sd:compatible}.}
\label{figure:configurations:6:point}
\end{figure}

Let \(2 \leq k \leq m\) be the smallest index such that \(\mathcal{S}_k\)
is not compatible. Such an index must exist in view of \(\mathcal{S}_m = \mathcal{S}_D\)
and our assumption that \(\mathcal{S}_D\) is not compatible.
The split system \(\mathcal{S}_{k-1}\), however, is compatible and there exists
an \(X\)-tree \(\mathcal{T}\) such that \(\mathcal{S}_{k-1} = \mathcal{S}_{\mathcal{T}}\).
Let \((x,y) = (u_k,v_k)\). The fact that
\(\mathcal{S}_k = \mathcal{S}_{k-1} \cup \{S_k\}\) is not compatible
implies that there must exist a vertex \(w\) on the path in \(\mathcal{T}\)
from the vertex labeled by \(x\) to the vertex labeled by \(y\)
such that the split \(S'\), corresponding to some edge \(e\) in \(\mathcal{T}\)
which has one endpoint at \(w\), is not compatible with \(S_k = S_{x,y}\).
Let \(\{s,t\} \in \binom{X}{2}\) be such that \(S' = S_{s,t}\) or \(S' = S_{t,s}\).
The two possible configurations, depending on whether or not edge \(e\) lies on the
path from \(x\) to \(y\) in \(\mathcal{T}\), are depicted in Figure~\ref{figure:configurations:6:point},
where \(a \in X_{x,y}\) and \(b \in X - X_{x,y}\).
It can be checked that the configuration depicted in Figure~\ref{figure:configurations:6:point}(a)
implies that~(1) holds while the configuration
depicted in Figure~\ref{figure:configurations:6:point}(b)
implies that~(2) holds, as required.
\hfill\(\blacksquare\)\\

To illustrate that a characterization as in Theorem~\ref{theorem:six:point:sd:compatible}
with a \(k\)-point condition for \(k < 6\) is not possible, one can employ the
following distance \(D\) that was presented in \cite{kea-97a}:
\begin{center}
\begin{tabular}{c|rrrrrr}
$D$ & $a$ & $b$ & $s$ & $t$ & $x$ & $y$\\
\hline
$a$ & 0 & 6 & 5 & 4 & 13 & 14\\
$b$ & 6 & 0 & 2 & 3 & 12 & 11\\
$s$ & 5 & 2 & 0 & 1 & 8 & 9\\
$t$ & 4 & 3 & 1 & 0 & 10 & 7\\
$x$ & 13 & 12 & 8 & 10 & 0 & 15\\
$y$ & 14 & 11 & 9 & 7 & 15 & 0 
\end{tabular}
\end{center}
It is shown in \cite{kea-97a} that the restriction of \(D\) to any 5-element
subset of \(X=\{a,b,s,t,x,y\}\) yields a distance \(D'\) such that
\(\mathcal{S}_{D'}\) is compatible. The split system \(\mathcal{S}_D\),
however, contains the splits \(S_{x,y} = \{b,t,y\}|\{a,s,x\}\)
and \(S_{s,t} = \{b,s,x\}|\{a,t,y\}\) which are not compatible.

Before continuing we briefly mention some previous work that is related to
the midpath phylogeny mentioned above and that
is concerned with situations where the distance \(D\) on \(X\) is not known
and only the rankings of the elements in \(X\) generated by \(D\) are available.
The aim then is to find a weighted compatible
split system that represents these rankings.
Methods that follow this approach and which heavily rely on the 
midpath phylogeny are presented in \cite{gue-97a,kan-war-95a,kea-98a,kea-hay-99a}.
Moreover, in \cite{kea-97a} it is shown (see also \cite{kea-hay-99a}) that any compatible split system
that represents the rankings of the elements in \(X\) generated by \(D\) must contain
the splits of the midpath phylogeny. However, as shown in \cite{sha-far-01a},
in general, if the rankings can be represented at all, further splits must be added.
The decision problem of whether the rankings can be represented or not can be solved in
polynomial time if restricted to representations by compatible split systems in which every split 
is assigned the same positive weight \cite{kan-war-95a,kea-hay-99a}. 
However, if the splits in the compatible split system
can have arbitrary positive weights, then the problem is NP-hard \cite{sha-far-06a}.

\section{Orderly split systems}
\label{section:circular:split:systems}

Motivated by the main result of Bonnot et al. in \cite{bon-gue-96a}, 
we call a split system \(\mathcal{S}\) \emph{orderly} if for 
all \(q = \frac{p}{2} > 0\) and all non-negative weightings
\(\omega\) of the splits in \(\mathcal{S}\) there exists a 
non-negative weighting \(\omega'\) of the splits in \(\mathcal{S}\)
such that \(O_{p,q}(D_{(\mathcal{S},\omega)}) = D_{(\mathcal{S},\omega')}\).
The result of Bonnot et al. can then be restated as follows:
Every compatible split system is orderly. In
this section we show that another important class of split systems
also enjoys this property.

We begin by recalling that 
a split system \(\mathcal{S}\) on \(X\) is \emph{circular} if there
exists an ordering \(\theta = x_1,x_2,\dots,x_n\) of the elements in \(X\) such that,
for any split \(S \in \mathcal{S}\), there exist \(1 \leq i \leq j < n\) with
\[S = \{x_i,x_{i+1},\dots,x_j\}| X - \{x_i,x_{i+1},\dots,x_j\},\]
in which case we will say then that \(\mathcal{S}\) \emph{fits} on \(\theta\).
Circular split systems naturally appear in the 
context of the so-called \emph{split decomposition} of a 
distance (see \cite[Sec. 3]{ban-dre-92a}, where they are introduced), and have
applications in phylogenetics (see e.g. \cite{bry-mou-04a}).
Note that every compatible split system is circular but not vice versa~\cite{ban-dre-92a}.

From the definition it follows immediately that a maximal circular split
system on a set with \(n\) elements contains precisely \(\binom{n}{2}\) splits.
Hence, for fixed \(n\), just like for compatible split systems,
the maximal circular split systems are precisely the maximum
circular split systems.
Moreover, while in general the ordering \(\theta\) of the elements in \(X\) 
onto which a circular split system fits might not be unique, it follows
again immediately from the definition that any maximum circular split
system uniquely determines this ordering up to reversing and shifting it.

Now, a distance \(D\) on \(X\) is \emph{circular} if there exists a circular
split system \(\mathcal{S}\) with a non-negative weighting \(\omega\) such that
\(D=D_{(\mathcal{S},\omega)}\). It is 
shown in~\cite{che-fic-98a} (see also \cite{chr-far-96a}) that a distance \(D\) on \(X\)
is circular if and only if there exists an ordering \(\theta = x_1,x_2,\dots,x_n\) of the
elements in \(X\) such that
\begin{equation}
\label{equ:kalmanson:cond}
\max (D(x_i,x_j)+D(x_k,x_l),D(x_i,x_l)+D(x_j,x_k)) \leq D(x_i,x_k) + D(x_j,x_l)
\end{equation}
holds for all \(1 \leq i < j < k < l \leq n\) so that, in particular,
circular distances are equivalent to so-called \emph{Kalmanson distances}~\cite{kal-75a}.
Note that an ordering \(\theta\) of the elements in \(X\)
satisfies condition~(\ref{equ:kalmanson:cond}) for a circular distance \(D\) if and only if
the necessarily unique circular split system with strictly positive weighting that generates~\(D\) fits on \(\theta\).
We now make a useful observation concerning circular split systems.

\begin{lemma}
\label{lem:split:system:sd:circular}
Let \((\mathcal{S},\omega)\) be a circular split system on \(X\) with non-negative weighting~\(\omega\)
such that \(\mathcal{S}\) fits on the ordering \(\theta=x_1,x_2,\dots,x_n\) of the elements in~\(X\).
Then, for \(D=D_{(\mathcal{S},\omega)}\), the split system \(\mathcal{S}_D\) is circular and fits on~\(\theta\).
\end{lemma}

\emph{Proof:}
If \(\mathcal{S}_D = \emptyset\) then \(\mathcal{S}_D\) is circular and it fits on \(\theta\).
So assume that \(\mathcal{S}_D \neq \emptyset\) and consider an arbitrary split \(S \in \mathcal{S}_D\).
By the definition of \(\mathcal{S}_D\) there exist \(u,v \in X\) with \(u \neq v\) and
\(S=X_{u,v}|X - X_{u,v}\).
Note that we must have \(u \in X_{u,v}\) and \(v \in X - X_{u,v}\).
Assume for a contradiction that, even after possibly shifting~\(\theta\),
the elements in \(X_{u,v}\) do not form an interval of consecutive elements in
\(\theta\). This implies that there exist \(u' \in X_{u,v}\) with
\(u' \neq u\) and \(v' \in X - X_{u,v}\) with \(v' \neq v\) such
that, after possibly shifting and/or reversing \(\theta\),
the restriction of \(\theta\) to \(\{u,v,u',v'\}\) is \(u,v',u',v\).
Then, in view of the definition of \(X_{u,v}\), we must have
\[D(u,u') < D(v,u') \ \text{and} \ D(v,v') \leq D(u,v').\]
But this implies
\[D(u,u') + D(v,v') < D((v,u') + D(u,v'),\]
contradicting condition~(\ref{equ:kalmanson:cond}).
\hfill\(\blacksquare\)

As a consequence of Lemma~\ref{lem:split:system:sd:circular}
we immediately obtain the main result of this section:

\begin{theorem}
\label{theo:circ:split:system:orderly}
Every maximum circular split system is orderly.
\end{theorem}

\emph{Proof:}
Let \(\mathcal{S}\) be a maximum circular split system together with a
non-negative weighting \(\omega\) and put \(D = D_{(\mathcal{S},\omega)}\).
In view of \(q = \frac{p}{2}\), Equation~(\ref{equ:general:formula:od:splits}) can be written as
\[O = \sum_{\substack{(u,v) \in X \times X, u \neq v\\ \emptyset \subsetneq X_{u,v} \subsetneq X}} \frac{p}{2} \cdot D_{S_{u,v}} = \sum_{S \in \mathcal{S}_D} \omega'(S) \cdot D_S,\]
where, for each \(S \in \mathcal{S}_D\), the weight \(\omega'(S)\) is a certain non-negative integer multiple of~\(\frac{p}{2}\).
By Lemma~\ref{lem:split:system:sd:circular}, \(\mathcal{S}_D\) fits onto the unique ordering
\(\theta\) of the elements in \(X\) onto which the maximum circular split system \(\mathcal{S}\) fits.
Hence, we have \(\mathcal{S}_D \subseteq \mathcal{S}\), as required.
\hfill\(\blacksquare\)

\begin{cor}
\label{cor:circ:dist:circ:orddist}
For \(q = \frac{p}{2}\), the order distance \(O_{p,q}(D)\)
of a circular distance \(D\) is always circular.
\end{cor} 

Note that in Theorem~\ref{theo:circ:split:system:orderly} we assume 
that the circular split system is
maximum. To illustrate that, in contrast to compatible split systems,
we cannot remove this assumption, 
consider, for example, the non-maximum circular split system
\[\mathcal{S} = \{\{b\}|\{a,c,d\},\{a,b\}|\{c,d\},\{a,d\}|\{b,c\}\}\]
on \(X=\{a,b,c,d\}\) and the weighting \(\omega\) that assigns weight~1 to every split
in~\(\mathcal{S}\). This yields the order distance \(O=O_{2,1}(D)\)
associated to \(D = D_{(\mathcal{S},\omega)}\) with
\begin{align*}
&O(a,b)=O(a,c)=O(b,c)=8,\\
&O(a,d)=O(c,d)=4 \ \text{and} \ O(b,d)=10.
\end{align*}
which is generated as \(O=D_{(\mathcal{S}',\omega')}\)
by the weighted circular split system \(\mathcal{S}' \supsetneq \mathcal{S}\) with
\begin{align*}
&\omega'(\{a,b\}|\{c,d\}) = \omega'(\{a,d\}|\{b,c\}) = 3,\\
&\omega'(\{a\}|\{b,c,d\}) = \omega'(\{c\}|\{a,b,d\}) = 1 \ \text{and} \ \omega'(\{b\}|\{a,c,d\}) = 4.
\end{align*}
So, in general, if \(D\) is generated by a non-maximum circular split system \(\mathcal{S}\) on \(X\)
the order distance associated to \(D\) may be generated only by a proper superset of \(\mathcal{S}\).

\section{Linearly independent split systems}
\label{section:max:lin:independent}

Circular split systems \(\mathcal{S}\) on \(X\) are 
examples of \emph{linearly independent} split systems as introduced in~\cite{bry-dre-07a},
that is, the set of split distances \(\{D_S : S \in \mathcal{S}\}\) arising from $\mathcal S$
is linearly independent when viewed as elements of the vector space of all
symmetric bivariate maps \(D : X \times X \rightarrow \mathbb{R}\) with
\(D(x,x)=0\) for all \(x \in X\). 
Note that the dimension of this vector space clearly is \(\binom{n}{2}\) and,
in view of the fact that there exist linearly independent split systems of size \(\binom{n}{2}\)
(e.g. maximum circular split systems), this implies that a linearly
independent split system \(\mathcal{S}\) on \(X\) is maximal
with respect to set inclusion if and only if \(\mathcal{S}\) has maximum size \(\binom{n}{2}\).
Linearly independent split systems therefore provide a natural
generalization of circular split systems, and in this section
we explore to what extent Theorem~\ref{theo:circ:split:system:orderly}
can be generalized to maximum linearly independent split systems
that are not circular. 

In the following we always
assume~\(q=\frac{p}{2}\) and, to avoid fractions in computations, put~\(p=2\).
The technical lemma we state next provides a useful link between the combinatorial structure
of a linearly independent split system and the order distances that it generates.
To describe this link, we call a split system \(\mathcal{S}\)
on a set \(X\) with \(n \geq 4\) elements \emph{closed} if for any two
incompatible splits \(A_1|B_1\) and \(A_2|B_2\) in \(\mathcal{S}\) at least one of the
following holds:
\begin{itemize}
\item[(a)]
\(\mathcal{S}\) also contains the splits \(A_1 \cap A_2 |X-(A_1 \cap A_2)\), \(B_1 \cap A_2|X-(B_1 \cap A_2)\),
\(A_1 \cap B_2 |X-(A_1 \cap B_2)\) and \(B_1 \cap B_2|X-(B_1 \cap B_2)\).
\item[(b)]
\(|A_1 \cap A_2| \cdot |B_1 \cap B_2| = |A_1 \cap B_2| \cdot |B_1 \cap A_2|\)
and \(\mathcal{S}\) also contains the split \((A_1 \cap A_2) \cup (B_1 \cap B_2) | (A_1 \cap B_2) \cup (B_1 \cap A_2)\).
\item[(c)]
\(|A_1 \cap A_2| \cdot |B_1 \cap B_2| > |A_1 \cap B_2| \cdot |B_1 \cap A_2|\)
and \(\mathcal{S}\) also contains the splits \((A_1 \cap A_2) \cup (B_1 \cap B_2) | (A_1 \cap B_2) \cup (B_1 \cap A_2)\),
\(A_1 \cap A_2 |X-(A_1 \cap A_2)\) and \(B_1 \cap B_2|X-(B_1 \cap B_2)\).
\item[(d)]
\(|A_1 \cap A_2| \cdot |B_1 \cap B_2| < |A_1 \cap B_2| \cdot |B_1 \cap A_2|\)
and \(\mathcal{S}\) also contains the splits \((A_1 \cap A_2) \cup (B_1 \cap B_2) | (A_1 \cap B_2) \cup (B_1 \cap A_2)\),
\(B_1 \cap A_2|X-(B_1 \cap A_2\) and \(A_1 \cap B_2 |X-(A_1 \cap B_2)\).
\end{itemize}

\begin{lemma}
\label{lemma:supporting:2:incompatible:splits:closed}
Let \(X\) be a set with \(n \geq 4\) elements and \(\mathcal{S}\) a linearly
independent split system on~\(X\). If \(\mathcal{S}\) is orderly then \(\mathcal{S}\) is closed.
\end{lemma}

\emph{Proof:}
If \(\mathcal{S}\) is compatible then it is trivially closed.
So assume that \(\mathcal{S}\) is orderly and contains two incompatible splits
\(S_1=A_1|B_1\) and \(S_2=A_2|B_2\).
Let~\(\omega\) be the weighting of~\(\mathcal{S}\) with \(\omega(S_1)=\omega(S_2)=2\) and
\(\omega(S) = 0\) for all other \(S \in \mathcal{S}\). We consider the order distance
\(O=O(D)\) with \(D=D_{(\mathcal{S},\omega)}\). 
Then, putting \(n_1 = |A_1 \cap A_2|\), \(n_2 = |B_1 \cap A_2|\), 
\(n_3 = |A_1 \cap B_2|\) and \(n_4 = |B_1 \cap B_2|\), we have,
for all \(x_1 \in A_1 \cap A_2\), \(x_2 \in B_1 \cap A_2\), \(x_3 \in A_1 \cap B_2\)
and \(x_4 \in B_1 \cap B_2\),
\begin{align*}
O(x_1,x_2) &= n_1 n_4 + 2(n_1 n_2 + n_3 n_4) + n_2 n_3,\\
O(x_1,x_3) &= n_1 n_4 + 2(n_1 n_3 + n_2 n_4) + n_2 n_3,\\
O(x_1,x_4) &= 2(n_1 n_4 + n_1 n_2 + n_3 n_4 + n_1 n_3 + n_2 n_4),\\
O(x_2,x_3) &= 2(n_2 n_3 + n_1 n_2 + n_3 n_4 + n_1 n_3 + n_2 n_4),\\
O(x_2,x_4) &= n_1 n_4 + 2(n_1 n_3 + n_2 n_4) + n_2 n_3,\\
O(x_3,x_4) &= n_1 n_4 + 2(n_1 n_2 + n_3 n_4) + n_2 n_3
\end{align*}
and, for all other \(u,v \in X\), we have \(O(u,v) = 0\).
This implies that, for any non-negative weighting \(\omega'\) of the splits in \(\mathcal{S}(X)\) with
\[O = \sum_{S \in \mathcal{S}(X)} \omega'(S) \cdot D_S,\]
only the splits \(S\) in
\begin{align*}
\mathcal{S}^* = \{&A_1 \cap A_2 |X-(A_1 \cap A_2), B_1 \cap A_2|X-(B_1 \cap A_2),\\
                  &A_1 \cap B_2 |X-(A_1 \cap B_2), B_1 \cap B_2|X-(B_1 \cap B_2),\\
                  &A_1|B_1, A_2|B_2, (A_1 \cap A_2) \cup (B_1 \cap B_2) | (A_1 \cap B_2) \cup (B_1 \cap A_2)\}
\end{align*}
can have a weight \(\omega'(S) > 0\). Note that \(\mathcal{S}^*\) is not linearly independent but
every 6-element subset of \(\mathcal{S}^*\) is. Now, 
\[O = \sum_{S \in \mathcal{S}^*} \omega'(S) \cdot D_S,\]
can be viewed as a system of linear equations for the weights \(\omega'(S)\).
Assuming without loss of generality that \(n_2 n_3 \leq n_1 n_4\),
it can be checked that there are only the following two solutions~(i) and~(ii)
of this system with \(\omega(S) \geq 0\) for all \(S \in \mathcal{S}^*\)
and \(|\{S \in \mathcal{S}^* : \omega'(S) > 0\}| \leq 6\):
\begin{itemize}
\item[(i)]
\begin{align*}
  &\omega'(B_1 \cap A_2|X-(B_1 \cap A_2)) = \omega'(A_1 \cap B_2|X-(A_1 \cap B_2)) = n_2 n_3,\\
  &\omega'(A_1 \cap A_2|X-(A_1 \cap A_2)) = \omega'(B_1 \cap B_2|X-(B_1 \cap B_2)) = n_1 n_4,\\
  &\omega'(A_1|B_1) = 2(n_1 n_2 + n_3 n_4),\\
  &\omega'(A_2|B_2) = 2(n_2 n_4  + n_1 n_3) \ \text{and}\\
  &\omega'((A_1 \cap A_2) \cup (B_1 \cap B_2) | (A_1 \cap B_2) \cup (B_1 \cap A_2)) = 0
\end{align*}
\item[(ii)]
\begin{align*}
  &\omega'(B_1 \cap A_2|X-(B_1 \cap A_2)) = \omega'(A_1 \cap B_2|X-(A_1 \cap B_2)) = 0,\\
  &\omega'(A_1 \cap A_2|X-(A_1 \cap A_2)) = \omega'(B_1 \cap B_2|X-(B_1 \cap B_2)) = n_1 n_4 - n_2 n_3,\\
  &\omega'(A_1|B_1) = 2(n_1 n_2 + n_3 n_4) + n_2 n_3\\
  &\omega'(A_2|B_2) = 2(n_2 n_4 + n_1 n_3) + n_2 n_3 \ \text{and}\\
  &\omega'((A_1 \cap A_2) \cup (B_1 \cap B_2) | (A_1 \cap B_2) \cup (B_1 \cap A_2)) = n_2 n_3
\end{align*}
\end{itemize}
But this implies, in view of the assumption that \(O=D_{(\mathcal{S},\omega')}\) for some
non-negative weighting~\(\omega'\) of~\(\mathcal{S}\), that 
\(\mathcal{S}\) must be closed, as required.
\hfill\(\blacksquare\)

Now we use Lemma~\ref{lemma:supporting:2:incompatible:splits:closed} to
obtain the following property of non-circular maximum linearly independent
split systems on sets with 5~elements.

\begin{lemma}
\label{lemma:non:fitting:same:class:5:points}
Let \(X=\{a,b,c,d,e\}\) be a set with 5~elements. Then every maximum linearly independent split system
\(\mathcal{S}\) on \(X\) that is not circular is not orderly.
\end{lemma}

\emph{Proof:}
Let~\(\mathcal{S}\) be a maximum linearly independent split system on \(X\) that is not circular
and assume for a contradiction that \(\mathcal{S}\) is orderly and, therefore, by
Lemma~\ref{lemma:supporting:2:incompatible:splits:closed}, closed.
Since \(\mathcal{S}\) is a maximum linearly independent split system, we have \(|\mathcal{S}|=10\).
Thus, as any compatible split system on \(X\) contains at most 7~splits, \(\mathcal{S}\) must
contain two incompatible splits \(S_1\) and \(S_2\). Relabeling the elements in \(X\), if necessary,
we assume without loss of generality that \(S_1 = \{a,b\}|\{c,d,e\}\) and \(S_2 = \{b,c\}|\{a,d,e\}\).
Then, since \(\mathcal{S}\) is closed,
it must also contain the splits \(\{b\}|\{a,c,d,e\}\) and \(\{d,e\}|\{a,b,c\}\).
Moreover, in view of \(|\mathcal{S}| = 10\), \(\mathcal{S}\) must contain an additional
split \(S_3 = \{x,y\}|X - \{x,y\}\) for a 2-element subset \(\{x,y\} \subseteq X\) with
\(\{x,y\} \not \in \{\{a,b\},\{b,c\},\{d,e\}\}\). We consider three cases.

\textsl{Case 1}: \(S_3 = \{a,c\}|\{b,d,e\}\).
Then, in view of the fact that \(\mathcal{S}\) is closed, 
\(S_1,S_3 \in \mathcal{S}\) implies that \(\{a\}|\{b,c,d,e\} \in \mathcal{S}\) and,
similarly, \(S_2,S_3 \in \mathcal{S}\) implies that \(\{c\}|\{a,b,d,e\} \in \mathcal{S}\).
But then it can be checked that \(\mathcal{S}\) is not a linearly independent
split system, a contradiction.

\textsl{Case 2}: \(S_3 = \{b,e\}|\{a,c,d\}\). (Note that the case \(S_3 = \{b,d\}|\{a,c,e\}\) is symmetric.)
Then, in view of the fact that \(\mathcal{S}\) is closed,
\(\{d,e\}|\{a,b,c\},S_3 \in \mathcal{S}\) implies \(\{a,c\}|\{b,d,e\} \in \mathcal{S}\).
From this we obtain a contradiction as in Case~1.

\textsl{Case 3}: \(S_3 = \{c,d\}|\{a,b,e\}\). (Note that the cases \(S_3 = \{c,e\}|\{a,b,d\}\),
\(S_3 = \{a,d\}|\{b,c,e\}\) and \(S_3 = \{a,e\}|\{b,c,d\}\) are symmetric.)
Then, using again that \(\mathcal{S}\) is closed,
\(S_2,S_3 \in \mathcal{S}\) implies \(\{c\}|\{a,b,d,e\},\{a,e\}|\{b,c,d\} \in \mathcal{S}\).
Three further applications of the definition of closedness each yield
that the split \(\{z\}|X-\{z\}\) is also contained in \(\mathcal{S}\) for \(z \in \{a,d,e\}\).
But this implies that \(\mathcal{S}\) is a maximum circular split system on \(X\),
a contradiction.
\hfill\(\blacksquare\)

Lemma~\ref{lemma:non:fitting:same:class:5:points} together
with Theorem~\ref{theo:circ:split:system:orderly} yields the following
characterization of orderly split systems
amongst all maximum linearly independent split systems
on sets with 5~elements:

\begin{prop}
\label{prop:non:fitting:same:class:5:points}
A maximum linearly independent split system \(\mathcal{S}\) on a set \(X\)
with 5~elements is orderly if and only if it is circular.
\end{prop}

It can be checked that maximum circular split systems on sets with 4~elements
cannot be characterized as in Proposition~\ref{prop:non:fitting:same:class:5:points}.

\section{Flat split systems}
\label{section:flat:split:systems}

In the last section we showed that a maximum linearly independent 
split system~\(\mathcal{S}\) on a set \(X\) with 5~elements is 
circular if and only if it is orderly. We would like to know if this can be extended 
to all values of $n \geq 5$, but have not been able to find a 
proof (or counter-example). However, in this section we show that the result can be extended to 
a certain subclass of linearly independent 
split systems called \emph{flat} split systems (see e.g. \cite{spi-ngu-12a}) that 
includes all circular split systems, and that was
first introduced in \cite{bry-dre-07a} under the name of \emph{pseudoaffine} split systems.
Just like circular split systems, flat split systems have found applications
in phylogenetics (see e.g. \cite{bal-spi-14a}).

To define this class of split systems we first associate
to any ordering \(\pi = (x_1,x_2,\dots,x_n)\) of \(X\), 
and any \(k \in \{1,2,\dots,n-1\}\) the ordering 
\[
\pi(k) = (x_1,\dots,x_{k-1},x_{k+1},x_k,x_{k+2}\dots,x_n),
\]
that is, the ordering obtained by swapping the 
elements \(x_k\) and \(x_{k+1}\) in $\pi$. 
In addition, we associate with this swap the 
subset \(sw(\pi,k) := \{x_k,x_{k+1}\}\) and
the the split 
\[
S(\pi,k) = \{x_1,\dots,x_k\}|\{x_{k+1},\dots,x_n\}. 
\]
Then, putting \(m = \binom{n}{2}\),
a pair \((\pi,\kappa)\) consisting of an ordering
\(\pi\) of \(X\) and a sequence 
\(\kappa = (k_1,\dots,k_m) \in \{1,2,\dots,n-1\}^m\)
is \emph{allowable} if
the sequence \(\pi_0,\pi_1,\dots,\pi_m\) of orderings
of \(X\) defined by putting
\(\pi_0 = \pi\) and \(\pi_i = \pi_{i-1}(k_i)\),
\(1 \leq i \leq m\), has the property that
any two elements in \(X\) swap their positions
precisely once, that is, 
\(sw(\pi_{i-1},k_i) \neq sw(\pi_{j-1},k_j)\) holds
for all \(1 \leq i < j \leq m\).
Note that such a sequence of orderings is commonly called
a \emph{simple allowable sequence} \cite{fel-04b}.

Now, a split system \(\mathcal{S} \subseteq \mathcal{S}(X)\) is called \emph{flat}
if there exists an allowable pair \((\pi,\kappa)\) 
with 
\[\mathcal{S} \subseteq \mathcal{S}_{(\pi,\kappa)} = \{S(\pi_{i-1},k_i) : 1 \leq i \leq m\}.\]
Here we are only concerned with \emph{maximal} flat split systems
which, similarly to circular split systems, are precisely the flat
split systems with maximum size~\(\binom{n}{2}\).
Note that in \cite[Theorem 14]{bal-bry-17a} it is shown that
a maximum linearly independent split system \(\mathcal{S}\) on a set \(X\) with \(n \geq 2\) elements
is a maximum flat split system if and only if, for every 4-element subset \(Y \subseteq X\), the restriction \(\mathcal{S}_{|Y}\)
contains precisely 6~splits, where
the \emph{restriction} of a split system \(\mathcal{S}\) on a set \(X\) to a subset
\(Y \subseteq X\) is the split system
\[\mathcal{S}_{|Y} = \{A \cap Y|B \cap Y : A|B \in \mathcal{S}\}.\]

Our goal is to provide a characterization of orderly
split systems amongst all maximum flat split systems 
similar to Proposition~\ref{prop:non:fitting:same:class:5:points}.
The following technical lemma will be used to obtain this characterization.
A split system \(\mathcal{S}\) on \(X\) has the \emph{pairwise separation property}
if, for any two distinct elements \(x,y \in X\),
there exist subsets \(A\) and \(B\) of \(X - \{x,y\}\) such that
\(A \cup \{x,y\}|B\), \(A \cup \{x\}|B \cup \{y\}\), \(A \cup \{y\}|B \cup \{x\}\) and
\(A|B \cup \{x,y\}\), if they form splits of \(X\), are contained in \(\mathcal{S}\). Note that every
circular split system satisfies the pairwise separation property~\cite{bry-dre-07a}. 

\begin{lemma}
\label{supporting:lemma:remove:one:element:pairwise:separation:property}
Let \(\mathcal{S}\) be a maximum linearly independent split system on a set \(X\) with
\(n \geq 3\) elements that satisfies the pairwise separation property. Then, for any \(y \in X\),
the restriction of \(\mathcal{S}\) to \(X - \{y\}\) is a maximum linearly independent split
system that satisfies the pairwise separation property.
\end{lemma}

\emph{Proof:}
Fix an arbitrary \(y \in X\). For every \(c \in X - \{y\}\) let
\(A_c\) and \(B_c\) denote the two subsets of \(X - \{y,c\}\) that
must exist for the pair \(\{y,c\}\) according to the pairwise separation
property. Consider the set of splits
\[\mathcal{S}_{\leftrightarrow} = \{A_c|X-(A_c \cup \{y\}) : c \in X - \{y\}\} \cup \{B_c|X-(B_c \cup \{y\}) : c \in X - \{y\}\}\]
on the set \(X - \{y\}\). Note that in the context of this proof it will
be convenient to consider \(\emptyset|X-\{y\}\) as a split of \(X\)
that may be contained in \(\mathcal{S}_{\leftrightarrow}\).

We claim that \(|\mathcal{S}_{\leftrightarrow}| \geq n-1\). To see this, consider the
graph \(G_{\leftrightarrow}\) with vertex set \(\mathcal{S}_{\leftrightarrow}\) in which
there is an edge between two distinct splits \(S\) and \(S'\) if there exists
a \(c \in X - \{y\}\) such that \(S=A_c|X-(A_c \cup \{y\})\) and \(S'=B_c|X-(B_c \cup \{y\})\).
Note that \(G_{\leftrightarrow}\) has the following properties:
\begin{itemize}
\item[(i)]
\(G_{\leftrightarrow}\) has \(n-1\) edges.
\item[(ii)]
Every cycle in \(G_{\leftrightarrow}\) has length \(n-1\).
\end{itemize}
Property~(i) follows directly from the definition
of \(G_{\leftrightarrow}\). Property~(ii) follows from the fact that,
in order to arrive along a cycle in \(G_{\leftrightarrow}\) at the same
split \(S = A|B\) again, all elements in \(A \cup B\) must move from
one side of the split to the other. Since \(A \cup B = X - \{y\}\), this
requires \(n-1\) moves.
Now, note that properties~(i) and~(ii) imply that \(G_{\leftrightarrow}\) must have at least
\(n-1\) vertices. Hence, we have \(|\mathcal{S}_{\leftrightarrow}| \geq n-1\), as claimed.

Next note that \(|\mathcal{S}_{\leftrightarrow}| \geq n-1\) implies that,
when restricting \(\mathcal{S}\) to \(X - \{y\}\),
we obtain at most \(\binom{n}{2} - (n-1)\) splits of \(X - \{y\}\) in view
of the fact that, for every \(S' \in \mathcal{S}_{\leftrightarrow}\), there
are, according to the pairwise separation property, at least two distinct
splits in \(\mathcal{S}\) that restrict to \(S'\). On the other hand,
in view of the fact that the square matrix consisting of the \(\binom{n}{2}\)
column vectors formed by the distances \(D_S\) for \(S \in \mathcal{S}\) has
full rank, the restriction of this matrix to the rows associated to
the pairs of distinct elements \(x,x' \in X - \{y\}\) must have rank
\(\binom{n}{2} - (n-1)\). But this implies that the restriction of \(\mathcal{S}\) to \(X - \{y\}\)
must contain at least \(\binom{n}{2} - (n-1)\) splits.
Thus \(|\mathcal{S}_{|X - \{y\}}| = \binom{n}{2} - (n-1)\), implying that
\(\mathcal{S}_{|X - \{y\}}\) is a maximum linearly independent split system on \(X-\{y\}\).
That \(\mathcal{S}_{|X - \{y\}}\) also satisfies the pairwise separation property
follows immediately from the definition of this property.
\hfill\(\blacksquare\)

\begin{cor}
\label{cor:flat:is:pairwise:separation}
A maximum linearly independent split system \(\mathcal{S}\) on a set \(X\) with \(n \geq 2\) elements
is maximum flat if and only if it satisfies the pairwise separation property.  
\end{cor}

\emph{Proof:}
First assume that \(\mathcal{S}\) is a maximum flat split system.
Then, by definition, there exists an allowable pair
\((\pi,\kappa)\) with \(\mathcal{S} = \mathcal{S}_{(\pi,\kappa)}\) and,
for any two distinct elements \(x,y \in X\), there must exist some
\(1 \leq i \leq \binom{n}{2}\) such that
\(\{x,y\} = sw(\pi_{i-1},k_i)\). Putting \(\pi_{i-1} = x_1,x_2,\dots,x_n\) and
\(k = k_i\), we obtain the two subsets \(A = \{x_1,x_2,\dots,x_{k-1}\}\) and \(B = \{x_{k+2},\dots,x_{n-1},x_n\}\) of \(X\)
for which it can be checked that \(A \cup \{x,y\}|B\), \(A \cup \{x\}|B \cup \{y\}\), \(A \cup \{y\}|B \cup \{x\}\) and
\(A|B \cup \{x,y\}\), if they form splits of \(X\),
are contained in \(\mathcal{S}_{(\pi,\kappa)}\) and, therefore, in~\(\mathcal{S}\), as required.

Next assume that \(\mathcal{S}\) is a maximum linearly independent split system
that satisfies the pairwise separation property.
Then, for \(n \in \{2,3,4\}\), \cite[Theorem~14]{bal-bry-17a} immediately implies that \(\mathcal{S}\)
is a maximum flat split system.
So, assume that \(n \geq 5\). Then, every 4-element subset \(Y \subseteq X\) can be obtained
by removing from \(X\) the elements in \(X - Y\) one by one. Hence, the restriction
\(\mathcal{S}_{|Y}\) is the last element of a sequence of restrictions, each to a subset
with one element less, and to each
such restriction Lemma~\ref{supporting:lemma:remove:one:element:pairwise:separation:property}
applies. Therefore, \(\mathcal{S}_{|Y}\) must be a linearly independent split system on
\(Y\) containing \(\binom{4}{2}=6\) splits. But this implies, again by \cite[Theorem~14]{bal-bry-17a},
that \(\mathcal{S}\) is maximum flat.
\hfill\(\blacksquare\)\\

We conclude this section with the above-mentioned characterization of orderly split
systems amongst all maximum flat split systems.

\begin{theorem}\label{flat-orderly}
\label{theorem:non:fitting:same:class:pairwise:separation:property}
Let \(X\) be a set with \(n \geq 5\)~elements and let \(\mathcal{S}\) be a
maximum flat split system on \(X\).
Then \(\mathcal{S}\) is orderly if and only if it is circular.
\end{theorem}

\emph{Proof:}
First assume that \(\mathcal{S}\) is a maximum circular split system.
Then, in view of Theorem~\ref{theo:circ:split:system:orderly},
\(\mathcal{S}\) is orderly.

Next assume that \(\mathcal{S}\) is a maximum flat split system that is not circular.
By Corollary~\ref{cor:flat:is:pairwise:separation}, \(\mathcal{S}\) satisfies
the pairwise separation property. We first show by induction on~\(n\) that
\begin{itemize}
\item[(e)]
\(\mathcal{S}\) contains two incompatible splits \(A_1|B_1\) and \(A_2|B_2\) 
such that the split \((A_1 \cap A_2) \cup (B_1 \cap B_2) | (A_1 \cap B_2) \cup (B_1 \cap A_2)\)
and at least one of the splits \(A_1 \cap A_2 |X-(A_1 \cap A_2)\) or \(B_1 \cap B_2|X-(B_1 \cap B_2)\)
are not contained in \(\mathcal{S}\).
\end{itemize}
The base case of the induction for \(n=5\) is established by directly checking the
following two split systems \(\mathcal{S}_1\)
and \(\mathcal{S}_2\) on the set \(X=\{a,b,c,d,e\}\) that are,
up to isomorphism, the only maximum flat split systems on \(X\) that are not circular:
\begin{align*}
\mathcal{S}_1 = \{&\{a\}|\{b,c,d,e\},\{b\}|\{a,c,d,e\},\{c\}|\{a,b,d,e\},\{d\}|\{a,b,c,e\},\\
                  &\{a,b\}|\{c,d,e\},\{b,c\}|\{a,d,e\},\{c,d\}|\{a,b,e\},\{a,d\}|\{b,c,e\},\\
                  &\{a,e\}|\{b,c,d\},\{b,e\}|\{a,c,d\}\}\\
\mathcal{S}_2 = \{&\{a\}|\{b,c,d,e\},\{b\}|\{a,c,d,e\},\{c\}|\{a,b,d,e\}\\
                  &\{a,b\}|\{c,d,e\},\{b,c\}|\{a,d,e\},\{c,d\}|\{a,b,e\},\{a,d\}|\{b,c,e\},\\
                  &\{a,c\}|\{b,d,e\},\{a,e\}|\{b,c,d\},\{b,e\}|\{a,c,d\}\}
\end{align*}
For \(n \geq 6\) select an element \(y \in X\) such that the restriction
\(\mathcal{S}' = \mathcal{S}_{|X-\{y\}}\) is not circular.
Since \(\mathcal{S}\) is not circular, such an element must exist.
By Lemma~\ref{supporting:lemma:remove:one:element:pairwise:separation:property}
and Corollary~\ref{cor:flat:is:pairwise:separation},
we have that \(\mathcal{S}'\) is a maximum flat split system on \(X-\{y\}\)
that is not circular. Thus, by induction, \(\mathcal{S}'\) satisfies~(e).
Then \(\mathcal{S}\) must contain two incompatible splits \(S_1\) and \(S_2\) such
their restriction to \(X-\{y\}\) satisfies~(e).
But then \(\mathcal{S}\) must satisfy~(e) for the two
incompatible splits \(S_1\) and \(S_2\) as well, finishing the inductive proof
that \(\mathcal{S}\) satisfies~(e).

To finish the proof of the theorem, it suffices to note that~(e)
implies that \(\mathcal{S}\) is not closed, which yields,
in view of Lemma~\ref{lemma:supporting:2:incompatible:splits:closed},
that \(\mathcal{S}\) is not orderly.
\hfill\(\blacksquare\)

\section{Computing the order distance}
\label{section:algorithms:computing:order:distance}

In this section we briefly look into algorithms for computing
the order distance \(O=O(D)=O_{p,q}(D)\) from an input distance \(D\) on a set
\(X\) with \(n\) elements. In~\cite{gue-98a} it is noted that
a run time in \(\mathcal{O}(n^4)\) can be achieved without any restrictions
on the input distance and the values of~\(p\) and~\(q\). Moreover,
clearly no algorithm for computing \(O(D)\) can 
run faster than the size of the output, that is,
we have a lower bound of \(\Omega(n^2)\).

As mentioned in Section~\ref{section:introduction},
the order distance \(O\) associated to a distance \(D\) on \(X\) can be viewed as a way
to quantify, for any two elements \(u,v \in X\), the differences between the rankings 
\(R_u\) and \(R_v\) of the elements in \(X\) generated by~\(D\) when sorting according to
non-decreasing distances from \(u\) and \(v\), respectively.
For non-generic distances \(D\) the rankings \(R_u\) and \(R_v\) are \emph{partial} in the sense that
ties between elements can occur when they have the same distance from \(u\) or \(v\).
In fact, it can be checked that the value \(O(u,v)\) coincides with
the so-called \emph{Kendall distance} \(K^{\pi}(R_u,R_v)\)
\emph{with penalty parameter}~\(\pi=\frac{q}{p}\) between the partial rankings \(R_u\) and \(R_v\)
that was introduced in~\cite{fag-kum-06a}. In~\cite{fag-kum-06a} the range of the
penalty parameter \(\pi\) is restricted to \(0 \leq \pi \leq 1\) and
it is also established that the Kendall distance \(K^{\pi}\) on partial rankings
of a fixed set \(X\) satisfies the triangle
inequality if and only if \(\frac{1}{2} \leq \pi \leq 1\).
In~\cite{ban-fer-09a} it is noted that, for any \(\pi \geq 0\), the
Kendall distance \(K^{\pi}(R_1,R_2)\) between two partial rankings \(R_1\) and \(R_2\)
of a set with \(n\) elements can be computed in \(\mathcal{O}(n \log n)\) time in a purely
comparison based model of computation and it is established that
in more powerful models of computation a run time in \(\mathcal{O}(n \log n / \log(\log n))\) can be
achieved (see also \cite{cha-pat-10a} for further related results).
Applying the algorithm from~\cite{ban-fer-09a} to each pair of elements
in \(X\) we therefore obtain:

\begin{prop}
\label{prop:run:time:general:input:distance}
The order distance \(O_{p,q}(D)\) of a distance \(D\) on a set with \(n\)
elements can be computed in \(\mathcal{O}(n^3 \log n)\) time.
\end{prop}

An immediate question is whether the gap to the lower bound \(\Omega(n^2)\) on the run
time can be further narrowed. In the following we will
show that this is the case for circular distances when we set \(q = \frac{p}{2}\). For the special
case of weighted compatible split systems a run time in \(\mathcal{O}(n^2 \log n)\)
follows immediately from the fact that, as established in~\cite{kea-97a},
the midpath phylogeny associated to an input distance \(D\) on a set with \(n\)
elements can be computed in \(\mathcal{O}(n^2 \log n)\) time in a purely
comparison based model of computation and in \(\mathcal{O}(n^2)\) 
in models of computation that allow for sorting in linear time.
We now show that a similar run-time can also be achieved for circular distances:

\begin{theorem}
\label{theorem:run:time:circular}
The order distance \(O(D) = O_{p,\frac{p}{2}}(D)\) of a circular distance \(D\)
on a set~\(X\) with~\(n\) elements can be computed in \(\mathcal{O}(n^2 \log n)\) time.
\end{theorem}

\emph{Proof:}
The first step in the computation of \(O(D)\) from \(D\)
is to obtain an ordering \(\theta = x_1,x_2,\dots,x_n\) of the elements in \(X\) such that,
\(D = D_{(\mathcal{S},\omega)}\) for a suitable weighting \(\omega\) of the circular split system \(\mathcal{S}\)
consisting of the splits \(S_{i,j}\) for which there exist \(1 \leq i \leq j < n\) with
\[S_{i,j} = \{x_i,x_{i+1},\dots,x_j\}| X - \{x_i,x_{i+1},\dots,x_j\}.\]
In view of the assumption that \(D\) is circular such an ordering must exist
and it can be computed in \(\mathcal{O}(n^2)\) time
with the algorithm presented in~\cite{chr-far-96a}.

Next note that, in view of Lemma~\ref{lem:split:system:sd:circular},
for all \(u,v \in X\) with \(D(u,v) > 0\), there exist
\(1 \leq i(u) \leq j(u) < n\) such that \(S_{i(u),j(u)} = S_{u,v}\)
as well as \(1 \leq i(v) \leq j(v) < n\) such that \(S_{i(v),j(v)} = S_{v,u}\).
Thus, using binary search, we can compute \(i(u)\), \(j(u)\), \(i(v)\) and \(j(v)\)
each in \(\mathcal{O}(\log n)\) time for fixed \(u,v \in X\). Doing this for all 
\(u,v \in X\) with \(D(u,v) > 0\), we obtain in \(\mathcal{O}(n^2 \log n)\) time
a weight \(\omega(S)\) for each split \(S \in \mathcal{S}\) such that
\(O=O(D) = D_{(\mathcal{S},\omega)}\).

Once we have the weighted split system \((\mathcal{S},\omega)\),
encoded as the ordering \(\theta\) and a non-negative
number representing \(\omega(S_{i,j})\) for all \(1 \leq i \leq j < n\),
it remains to compute \(O(x,y) = D_{(\mathcal{S},\omega)}(x,y)\)
for all \(x,y \in X\). It is known that this can done in \(\mathcal{O}(n^2)\) time
(see e.g.~\cite{mou-spi-12a} where this and related computational
problems on split systems are discussed).
\hfill\(\blacksquare\)

\section{Concluding remarks}
\label{section:comcluding remarks}

In this paper, we have shed some light on the relationship between
split systems and the order distances generated by them. 
A specific question that remains open is whether or not
the generalization of Lemma~\ref{lemma:non:fitting:same:class:5:points} to \(n \geq 6\)
is true, which would for example yield an interesting new characterization of maximum circular split systems
among all maximum linearly independent split systems.
Computational experiments that we have performed on a large number of randomly generated
maximum linearly independent split systems seem to indicate that, at least for \(n=6\), if counterexamples
exist they are very rare.

To explain another aspect of order distances and
maximum linearly independent split systems, note that, usually,
an \(\ell_1\)-distance~\(D\) on~\(X\) can be generated by many different
maximum linearly independent split systems on~\(X\). Therefore, for any 
\(\ell_1\)-distance $D$, one might hope to
at least always find \emph{some} such split system that would generate both \(D\) and \(O(D)\).
However, through an exhaustive search through the 34~isomorphism classes of maximum
linearly independent split systems on a set \(X\) with \(n=5\) elements
with a computer program we found that,
for every maximum linearly independent split system
\(\mathcal{S}\) on \(X\) that is not maximum flat, there exists a non-negative weighting
\(\omega\) such that for the distance \(D=D_{(\mathcal{S},\omega)}\) and its associated
order distance \(O(D)\) 
there is \emph{no} maximum
linearly independent split system
that generates both \(D\) and \(O(D)\).
In future work, it would be interesting to explore this further and see if
it yields a characterization of maximum flat split systems.

Another interesting direction could be to further explore
properties of the split system \(\mathcal{S}_D\). Note that from
Theorem~\ref{theo:upper:bound:size:sd} it follows that, in general,
\(\mathcal{S}_D\) will not be linearly independent because it
contains more than \(\binom{n}{2}\) splits. In contrast, if 
\(D\) is circular then, in view of Lemma~\ref{lem:split:system:sd:circular},
\(\mathcal{S}_D\) is circular too and, therefore, linearly independent.
So, a specific question one can ask is for which distances $D$
the split system \(\mathcal{S}_D\) is linearly independent?

Finally, in future work it could also be interesting to study variants of the
order distance that are obtained by employing, instead of the Kendall distance
with penalty parameter~\(\frac{q}{p}\), any of 
the other distance measures on partial rankings introduced in~\cite{fag-kum-06a}.

\subsection*{Acknowledgment}
The authors would like to thank Barbara Holland and Katharina Huber for interesting 
discussions on the ordering of treelike distances which led them to  
consider the connection between order distances and split systems.


\begin{thebibliography}{10}

\bibitem{bal-bry-17a}
M.~Balvociute, D.~Bryant, and A.~Spillner.
\newblock When can splits be drawn in the plane?
\newblock {\em SIAM Journal on Discrete Mathematics}, 31:839--856, 2017.

\bibitem{bal-spi-14a}
M.~Balvociute, A.~Spillner, and V.~Moulton.
\newblock Flat{NJ}: {A} novel network-based approach to visualize evolutionary
  and biogeographical relationships.
\newblock {\em Systematic Biology}, 63:383--396, 2014.

\bibitem{ban-dre-92a}
H.-J. Bandelt and A.~Dress.
\newblock A canonical decomposition theory for metrics on a finite set.
\newblock {\em Advances in Mathematics}, 92:47--105, 1992.

\bibitem{ban-fer-09a}
M.~Bansal and D.~Fern\'andez-Baca.
\newblock Computing distances between partial rankings.
\newblock {\em Information Processing Letters}, 109:238--241, 2009.

\bibitem{bon-gue-96a}
F.~Bonnot, A.~Gu\'enoche, and X.~Perrier.
\newblock Properties of an order distance associated to a tree distance.
\newblock In E.~Diday et~al., editor, {\em Ordinal and symbolic data analysis},
  pages 252--261. Springer, 1996.

\bibitem{bry-dre-07a}
D.~Bryant and A.~Dress.
\newblock Linearly independent split systems.
\newblock {\em European Journal of Combinatorics}, 28:1814--1831, 2007.

\bibitem{bry-mou-04a}
D.~Bryant and V.~Moulton.
\newblock {N}eighbor{N}et: An agglomerative method for the construction of
  phylogenetic networks.
\newblock {\em Molecular Biology and Evolution}, 21:255--265, 2004.

\bibitem{cha-pat-10a}
T.~Chan and M.~P\u{a}tra\c{s}cu.
\newblock Counting inversions, offline orthogonal range counting, and related
  problems.
\newblock In {\em Proc. 21st Annual ACM-SIAM Symposium on Discrete Algorithms},
  pages 161--173. SIAM, 2010.

\bibitem{che-fic-98a}
V.~Chepoi and B.~Fichet.
\newblock A note on circular decomposable metrics.
\newblock {\em Geometriae Dedicata}, 69:237--240, 1998.

\bibitem{chr-far-96a}
G.~Christopher, M.~Farach, and M.~Trick.
\newblock The structure of circular decomposable metrics.
\newblock In {\em Proc. 4th Annual European Symposium on Algorithms}, LNCS,
  pages 486--500. Springer, 1996.

\bibitem{dez-97}
M.~Deza and M.~Laurent.
\newblock {\em Geometry of Cuts and Metrics}.
\newblock Springer, 1997.

\bibitem{fag-kum-06a}
R.~Fagin, R.~Kumar, M.~Mahdian, D.~Sivakumar, and E.~Vee.
\newblock Comparing partial rankings.
\newblock {\em SIAM Journal on Discrete Mathematics}, 20:628--648, 2006.

\bibitem{fel-04b}
S.~Felsner.
\newblock {\em Geometric graphs and arrangements}.
\newblock Vieweg, 2004.

\bibitem{gue-97b}
A.~Gu\'enoche.
\newblock Order distance associated with a hierarchy.
\newblock {\em Journal of Classification}, 14:101--115, 1997.

\bibitem{gue-97a}
A.~Gu\'enoche.
\newblock Order distances in tree reconstruction.
\newblock In B.~Mirkin et~al., editor, {\em Mathematical Hierarchies and
  Biology}, pages 171--182. American Mathematical Society, 1997.

\bibitem{gue-98a}
A.~Gu\'enoche.
\newblock Ordinal properties of tree distances.
\newblock {\em Discrete Mathematics}, 192:103--117, 1998.

\bibitem{kal-75a}
K.~Kalmanson.
\newblock Edgeconvex circuits and the travelling salesman problem.
\newblock {\em Canadian Journal of Mathematics}, 27:1000--1010, 1975.

\bibitem{kan-war-95a}
S.~Kannan and T.~Warnow.
\newblock Tree reconstruction from partial orders.
\newblock {\em SIAM Journal on Computing}, 24:511--519, 1995.

\bibitem{kea-97a}
P.~Kearney.
\newblock A six-point condition for ordinal matrices.
\newblock {\em Journal of Computational Biology}, 4:143--156, 1997.

\bibitem{kea-98a}
P.~Kearney.
\newblock The ordinal quartet method.
\newblock In {\em Proc. 2nd Annual International Conference on Computational
  Molecular Biology}, pages 125--134. ACM, 1998.

\bibitem{kea-hay-99a}
P.~Kearney, R.~Hayward, and H.~Meijer.
\newblock Evolutionary trees and ordinal assertions.
\newblock {\em Algorithmica}, 25:196--221, 1999.

\bibitem{mou-spi-12a}
V.~Moulton and A.~Spillner.
\newblock Optimal algorithms for computing edge weights in planar split
  networks.
\newblock {\em Journal of Applied Mathematics and Computing}, 39:1--13, 2012.

\bibitem{sem-ste-03a}
C.~Semple and M.~Steel.
\newblock {\em Phylogenetics}.
\newblock Oxford University Press, 2003.

\bibitem{sha-far-01a}
R.~Shah and M.~Farach-Colton.
\newblock On the midpath tree conjuncture: a counter-example.
\newblock In {\em Proc. 12th Annual ACM-SIAM Symposium on Discrete Algorithms},
  pages 207--208. SIAM, 2001.

\bibitem{sha-far-06a}
R.~Shah and M.~Farach-Colton.
\newblock On the complexity of ordinal clustering.
\newblock {\em Journal of Classification}, 23:79--102, 2006.

\bibitem{spi-ngu-12a}
A.~Spillner, B.~Nguyen, and V.~Moulton.
\newblock Constructing and drawing regular planar split networks.
\newblock {\em IEEE/ACM Transactions on Computational Biology and
  Bioinformatics}, 9:395--407, 2012.

\end{thebibliography}
\end{document}